\documentclass[cits,a4paper]{PoS}
\usepackage{amsmath,amssymb}
\usepackage[tight,TABTOPCAP]{subfigure}
\usepackage{graphicx}
\usepackage{grffile}

\title{The hadronic vacuum polarization function with $O(a)$-improved Wilson fermions - an update}

\ShortTitle{The HVP with $O(a)$-improved Wilson fermions - an update}

\author{\vspace{-0.75cm}\phantom{a} \hfill \textnormal{MITP/16-004}\newline\newline Michele Della Morte$^1$, Gregorio
Herdoiza$^4$, \speaker{Hanno Horch}$^3$, Benjamin J\"ager$^{5}$, Harvey
Meyer$^3$, Hartmut Wittig$^{2,3}$\\
$^{1}$ CP3-Origins \& Danish IAS, University of Southern Denmark\\
		 Campusvej 55, DK-5230 Odense M, Denmark and IFIC (CSIC)\\
		 Calle Catedr\'{a}tico Jos\'{e} Beltr\'{a}n, 2. E-46980, Paterna, Spain\\
$^{2}$Helmholtz Institute Mainz, Johannes Gutenberg Universit\"at Mainz,
       55099 Mainz, Germany \\
 $^{3}$PRISMA Cluster of Excellence, Institut f\"ur Kernphysik, Johannes
	Gutenberg Universit\"at Mainz, 55099 Mainz, Germany\\
$^{4}$  Instituto de F\'isica Te\'orica UAM/CSIC and
Departamento de F\'isica Te\'orica,\\
Universidad Aut\'onoma de Madrid, Cantoblanco, E-28049
Madrid, Spain\\
$^{5}$ Department of Physics, College of Science, Swansea University, SA2 8PP, Swansea, UK\\
E-mail:\email{horch@kph.uni-mainz.de}, \email{dellamor@cp3-origins.net}
	   \email{gregorio.herdoiza@uam.es},
       \email{B.Jaeger@swansea.ac.uk},
       \email{meyerh@kph.uni-mainz.de}, \email{wittig@kph.uni-mainz.de}\\}

\abstract{We present an update of our lattice QCD study of the vacuum polarization function
using O$(a)$-improved $N_ {\rm f} =2$ Wilson fermions with increased statistics and a large set of
momenta. The resulting points are highly correlated and thus require a correlated fitting procedure.
We employ an extended frequentist method to estimate the systematic uncertainties due to the
momentum dependence and to the continuum and chiral extrapolations. We present preliminary results 
of the leading order hadronic contribution of the anomalous magnetic moment of the muon
$\left(a_\mu^{\mathrm{HLO}}\right)$ at the physical point for $u,d,s$ and $c$ valence quarks.}

\FullConference{The 33rd International Symposium on Lattice Field Theory\\
		14 -18 July 2015\\
		Kobe International Conference Center, Kobe, Japan*}

\begin{document}

\section{Introduction}
The anomalous magnetic moment of the muon is one of the most precisely measured quantities in
physics. However, for a number of years there has been a persistent $\sim 3.5\,\sigma$ discrepancy
between experimental measurements and the prediction from theory \cite{PDG2014},
\begin{align*}
a_\mu^{exp}&=116\,592\,091(54)(33) \cdot 10^{-11},\\		
a_\mu^{th} &=116\,591\,803(01)(42)(26)\cdot 10^{-11}.	
\end{align*}
The theoretical error is dominated by hadronic contributions. The lowest-order hadronic
contribution is estimated using a dispersion relation relying on experimental data, so a
determination from first principles using lattice QCD is desirable. This has lead to interest in the
lattice community and several groups have reported results \cite{aubinblum07, Boyle:2011hu,
DellaMorte:2011aa, Burger:2013jya}. For the determination of the hadronic vacuum polarization (HVP)
tensor on the lattice we use
\begin{align}
\Pi_{\mu\nu}(Q)&=Z_V\sum_{x} e^{iQx}\left<J^{(c)}_\mu(x)J^{(l)}_\nu(0)\right>,
\end{align} 
where $J^{(c)}_\mu(x),$ and $J^{(l)}_\nu(x)$ refer to conserved and local vector currents,
respectively, and $Z_V$ is the renormalization factor of the local current
\cite{DellaMorte:2005xgj}. The HVP is then given by
\begin{align}
\Pi_{\mu\nu}(Q)=\left(Q_\mu Q_\nu - \delta_{\mu\nu}Q^2\right)\Pi(Q^2).
\end{align}
To determine $a_\mu^{\mathrm{HLO}}$ the renormalized HVP,
$\widehat{\Pi}(Q^2)=4\pi^2(\Pi(Q^2)-\Pi(0))$, is inserted into the convolution integral
\cite{deRafael:1993za, Blum:2002ii}
\begin{align}
  a_\mu^{\rm HLO} & =  \left(\frac{\alpha}{\pi}\right)^2 \,
  	\int_{0}^{\infty} \frac{dQ^2}{Q^2} \, w(Q^2/m_\mu^2) \,\widehat{\Pi}(Q^2)
  	\label{convolution_integral},\\
  w(r) &= 16/\left(r^2 \, \left( 1 + \sqrt{1+4/r} \right)^4 \sqrt{1+4/r}\right),  
\end{align}
where the integrand in eq. (\ref{convolution_integral}) is dominated by the region around $Q^2 \sim
m_\mu^2$.

\section{Lattice setup and the extended frequentist method}
We use $O(a)-$improved Wilson fermions with two dynamical degenerate light quarks and
partially quenched strange and charm quarks. We use the ensembles generated within the
CLS effort listed in table \ref{clsensembles}. Twisted boundary conditions are applied to increase
the number of available momenta and to gain access to small momenta
\cite{Sachrajda:2004mi, Bedaque:2004ax, deDivitiis:2004kq}.

\begin{figure}[ht!]
\centering 
\includegraphics[scale=0.75]{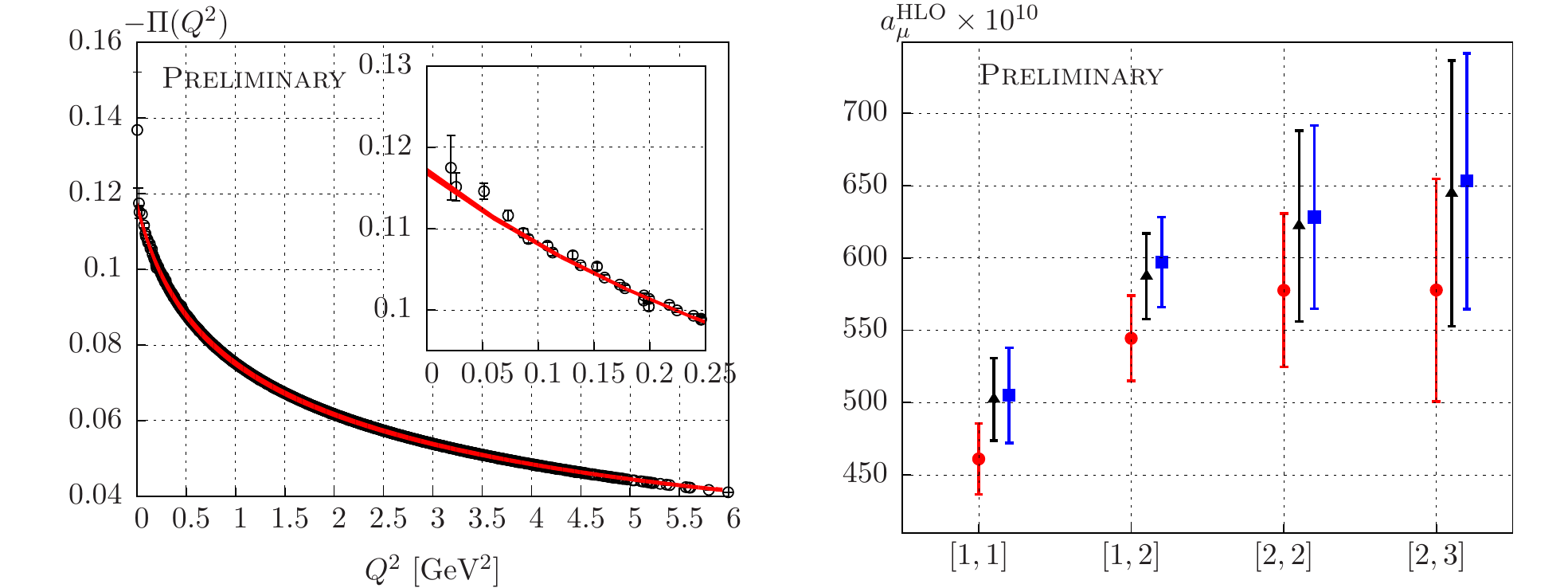}
\caption{{\bf Left}: The result for the HVP on G8, $m_\pi=$185 MeV,
$a=$0.0658 fm, is shown in black. We combine the results determined via the extended
frequentist method for the Pad\'e $[1,2]$, and a zoom into the small $Q^2$ region. We blow up
the errors of the result in the plot of the full $Q^2$ range to make the curve visible. {\bf Right}: We
tested different orders of the Pad\'{e} approximants for the fit functions and found that in fit
interval used in this study the $p$-value of the fits with Pad\'{e} order below [1,2] is low, while
including $[2,3]$ would only increase the error without adding new information.}
\label{g8_ud_pade12_ordertest}
\end{figure}

\begin{table}[ht!]
\begin{center}
\begin{tabular}{|c|c|c|c|c|c|c|}
\hline
Label	&	$L/a$	&	$\beta$	&	$am_\pi$	&	$m_\pi L$	&	$a$ [fm]	&	$m_\pi$ [MeV]\\
\hline
\hline
A3	&	32		&	5.20	&	0.1893(6)	&	6.1	&	0.0792(26)	&	472\\
A4	&	32		&	5.20	&	0.1459(7)	&	4.7	&	0.0792(26)	&	364\\
A5	&	32		&	5.20	&	0.1265(8)	&	4.0	&	0.0792(26)	&	315\\
B6	&	48		&	5.20	&	0.1073(7)	&	5.2	&	0.0792(26)	&	267\\
\hline
E5	&	32		&	5.30	&	0.1458(3)	&	4.7	&	0.0631(21)	&	456\\
F6 	&	48		&	5.30	&	0.1036(3)	&	5.0	&	0.0631(21)	&	324\\
F7	&	48		&	5.30	&	0.0885(3)	&	4.2	&	0.0631(21)	&	277\\
G8	&	64		&	5.30	&	0.0617(3)	&	3.9	&	0.0631(21)	&	193\\
\hline
N5	&	48		&	5.50	&	0.1086(2)	&	5.2	&	0.0499(19)	&	429\\
N6	&	48		&	5.50	&	0.0838(2)	&	4.0	&	0.0499(19)	&	331\\
O7	&	64		&	5.50	&	0.0660(1)	&	4.2	&	0.0499(19)	&	261\\
\hline
\end{tabular}
\caption[CLS ensembles]{The CLS ensembles used in this study. We use the determination of the
scale via $f_K$ \cite{Fritzsch:2012wq} and the masses determined in \cite{Capitani:2015sba}.}
\label{clsensembles}
\end{center}
\end{table}
The data are highly correlated among the $Q^2$ momenta, and the large number of data points often
lead to singular correlated covariance matrices. To avoid singularities, we randomly select
subsets of 30 and 40 points in the interval $0 < Q^2 < 4\,{\rm ~GeV}^2$. The data points
at larger $Q^2$ values are very precise but only represent a small contribution to the convolution
integral in eq. (\ref{convolution_integral}), so that we focus on data points in low $Q^2$ regime.
In order to determine the number of samples chosen, we compute the distribution in $a_\mu^{\rm HLO}$
with respect to the samples. We pick 1000 different samples of 30 and 40 data points, and the
variation in $a_\mu^{\rm HLO}$ is included in the systematic error estimate. The $Q^2$
dependence of the HVP is modelled by Pad\'{e} approximants \cite{DellaMorte:2011aa,
Aubin:2012me},
\begin{align}
\Pi^{\rm fit}_{1,2}(Q^2) & =\Pi(Q^2=0) -  Q^2\left(\frac{a_1}{b_1 + Q^2} + \frac{a_2}{b_2 +
Q^2}\right),\\
\Pi^{\rm fit}_{2,2}(Q^2) & =\Pi(Q^2=0) -  Q^2\left(a_0 + \frac{a_1}{b_1 + Q^2} + \frac{a_2}{b_2 +
Q^2}\right),
\label{fitfunctions}
\end{align}
where we impose that $a_{1,2} > 0$ and $b_{1,2} > 0$, and $\Pi(Q^2=0)$ is determined via an
extrapolation. This type of representation is known to converge to $\Pi(Q^2)$ \cite{Aubin:2012me}.
We impose a conservative cut, $0 < a_\mu^{\rm HLO} < 10^{-6}$, to avoid some isolated numerical
instabilities in the fits. As an example we show our results on our most chiral ensemble G8 in
figure \ref{g8_ud_pade12_ordertest}, which has the largest statistical uncertainties.\\
To determine the systematic error for a number of variations in the calculation, we use the
extended frequentist method \cite{PDG2006, Durr:2008zz}. For this procedure the central value is
given through the median of the central values of all variations, and the central 68\% of this
distribution associated with the systematic error. The statistical error is computed by the median
of each bootstrap sample for all variations. The statistical error is then given by the central 68\% of
the distribution of these medians. The analysis involves two steps, first a fit of the momentum
dependence of the HVP for each ensemble and then an extrapolation of $a_\mu^{\rm HLO}$ to the
physical point. In the implementation of the extended frequentist method for the second step, we
weight the distributions over the considered variations by the corresponding $p$-value of the fits.

\section{Extrapolation of $a_\mu^{\mathrm{HLO}}$ to the physical point}
\begin{figure}[ht!]
\centering 
\includegraphics[scale=0.65]{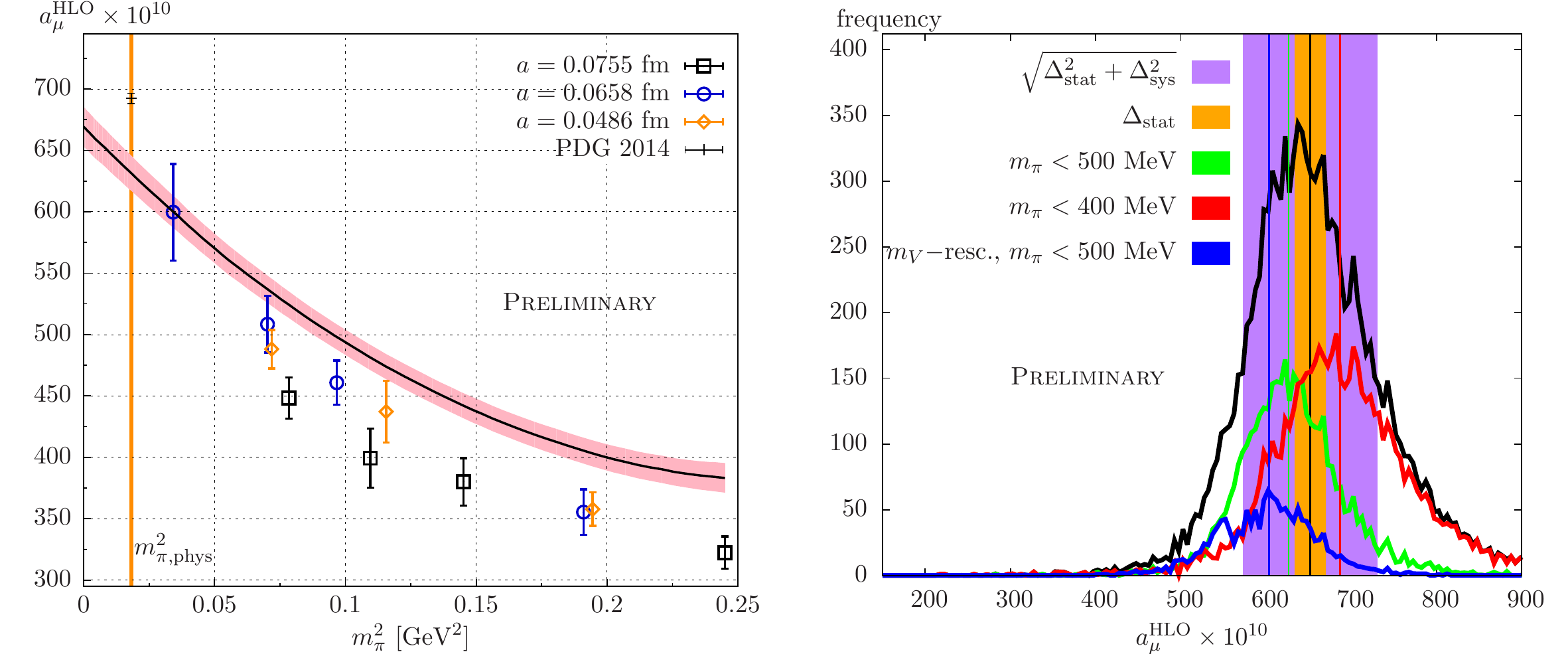}
\caption{Left: Example of an extrapolation to the physical point with $a_\mu^{\rm
HLO, A} (m_\pi^2, a)$ using all ensembles for $uds_Qc_Q$. The fit function is evaluated in the
continuum and thus appears above the data. The result from phenomenology is included as a
reference. Right: The histogram derived from all possible variations (black).
The statistical error is the orange band centered around the median shown in black of this histogram.
The purple band includes the total error. To study the different systematic effects we project out
different variations and build histograms with these projections of the data, the medians are shown
as vertical lines in the corresponding color for each projection, respectively. We show the
projections used to determine the contribution to the systematic error due to the chiral
extrapolation.}
\label{amu_chiral_extrapolation_plot}
\end{figure}

For the extrapolation to the physical point we fit the $m_\pi^2$-dependence and lattice artifacts of
$O(a)$ simultaneously using the following functions:
\begin{align}
 a_\mu^{\rm HLO, A} (m_\pi^2, a) & =  c_1 + c_2 m_\pi^2 + c_3 m_\pi^2\log(m_\pi^2) + c_4 a,\\
 a_\mu^{\rm HLO, B} (m_\pi^2, a) & =  c_1 + c_2 m_\pi^2 + c_3 m_\pi^4 + c_4 a,
 \end{align}
 where the fit ansatz $a_\mu^{\rm HLO, A} (m_\pi^2, a)$ is inspired by chiral perturbation theory,
 and $a_\mu^{\rm HLO, B} (m_\pi^2, a)$ is based on a more naive expansion in $m_\pi^2$. Following
 \cite{Renner:2012fa} we also consider a linear function, i.e. 
 \begin{align}
 a_\mu^{\rm HLO, C} (m_\pi^2, a) & =  c_1 + c_2 m_\pi^2 + c_3 a,
\end{align}
after rescaling the convolution function $w(r)$ in eq. (\ref{convolution_integral}) according to
\begin{align}
  w\left(\frac{Q^2}{m_\mu^2}\right) \longrightarrow w\left(\frac{Q^2}{m_\mu^2}\,
  \left(\frac{M_\rho^{\rm phys}}{M_V}\right)^2\right).
  \label{rescaling_of_integral_weight}
\end{align}
Here $M_V$ is the vector meson mass extracted from the vector correlation function, and the
additional physical input from the experimental $\rho$-meson mass, $M_\rho^{\rm phys}$, is inserted.
The rescaling in eq.~(\ref{rescaling_of_integral_weight}) provides an alternative for the chiral
extrapolation and results in a milder pion mass dependence for $a_\mu^{\rm HLO}$. For every fit
function we consider cuts on the contributing ensembles to the fit. For the fit functions of type
$a_\mu^{\rm HLO, A} (m_\pi^2, a)$ and $a_\mu^{\rm HLO, B} (m_\pi^2, a)$ we first consider all
ensembles and also impose cuts at $m_\pi<$400 MeV and $a<0.070$ fm. When $a<0.070$ fm is used we
switch off the term describing lattice artifacts, i.e. we set $c_4 = 0$, as we do not observe a
clear lattice spacing dependence for the data, see e.g. figure \ref{amu_chiral_extrapolation_plot}.
For the third ansatz, $a_\mu^{\rm HLO, C} (m_\pi^2, a)$, we consider all ensembles and the cut
$m_\pi<$400 MeV. To illustrate the method we show the result we obtain for the fit
function $a_\mu^{\rm HLO, A}$ on the left of figure \ref{amu_chiral_extrapolation_plot}, where we evaluated the fit function in the continuum limit,
which explains the vertical shift of the function with respect to the data.

\begin{figure}[ht!]
\centering
\includegraphics[scale=1]{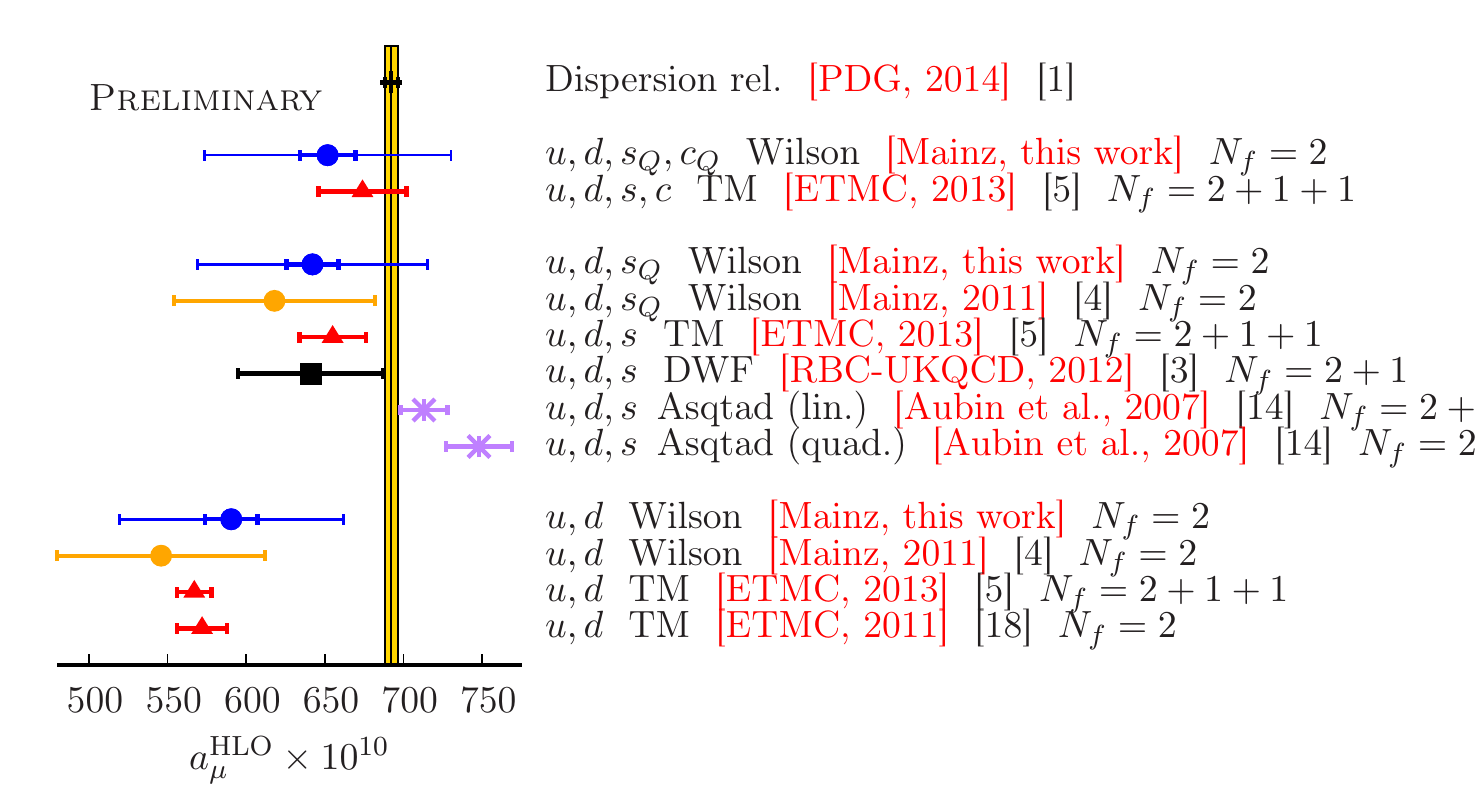}
\caption{We show our updated results in blue for $a_\mu^{\rm HLO}$ in comparison to the results of
various groups sorted by valence quark contribution. The inner error bar on the blue points
is the statistical error only, the outer error bar includes systematic errors summed in quadrature.}
\label{g-2_comparison}
\end{figure}

To study the dominant of systematic effects projections of all variations used
for the histogram are shown in black on the left of figure \ref{amu_chiral_extrapolation_plot}. For
each projection we compute the median, and the standard deviation of these medians gives a rough estimate for the
contribution to the total systematic error. To illustrate we show the projections for the effect
of the chiral extrapolation in red, blue, and green in the same figure. We have one subgroup for the
extrapolations based on $a_\mu^{\rm HLO, \{A,B,C\}} (m_\pi^2, a)$  using all ensembles, and one
subgroup for the fit functions $a_\mu^{\rm HLO, \{A,B\}} (m_\pi^2, a)$ with the cut $m_\pi < 400$
MeV. The third subgroup consists of the results obtained using $a_\mu^{\rm HLO, C} (m_\pi^2,
a)$. To probe the contribution due to lattice artifacts we use two subgroups. The first subgroup
consists of the fit functions $a_\mu^{\rm HLO, \{A,B,C\}} (m_\pi^2, a)$ where we use a term
proportional to lattice artifacts, i.e. $c_4\neq 0$ and $c_3\neq 0$, respectively. The second
consists of the fits where we set $c_4=0$ using only $a_\mu^{\rm HLO, \{A,B\}}(m_\pi^2, a)$. Other
sources of systematic error include the choice of Pad\'{e} approximant used to fit the HVP, as
well as the picking of samples of subsets of data points, in order to perform viable correlated
fits. We show the normalized results for the different contributions to the systematic error in
table \ref{systematiceffects}.
\begin{table}[ht!]
\begin{center}
\begin{tabular}{|c|c|c|c|}
\hline	
Label & $ud$ & $uds_Q$ & $uds_Qc_Q$ \\
\hline
 $\chi-$extrapolation 	 &	47\%	& 42\%	&	40\%\\
 Lattice artifacts 	  	 &	10\%	& 14\%	& 	15\%\\
 $Q^2$-sampling 	 	 &	31\%	& 31\%	&	34\%\\
 Pad\'{e} 	 			 &	11\%	& 12\%	& 	11\%\\
 Points/$Q^2-$sample 	 &	<1\%	& <1\%	& 	<1\%\\
\hline
\end{tabular}
\end{center}
\caption{We list the relative contribution of the sources of the systematic error for each flavor
combination separately.}
\label{systematiceffects}
\end{table}
We find that the uncertainty due to the chiral extrapolation dominates. The systematic effects
introduced due to picking samples of the HVP data is also a sizeable contribution. Lattice artifacts
and the choice of Pad\'{e} approximant for the fit to the HVP are of the same order, while the
number of points per $Q^2$-sample appears to be negligible.\newline
In figure \ref{g-2_comparison} we compare our preliminary results, shown in blue, to other
lattice groups sorted by valence quark contributions. The inner error bars show the statistical
error only, while the outer error bars include the systematic errors summed in quadrature. The
uncertainty on our preliminary results is dominated by a conservative estimate of the systematic
effects.
\section{Conclusions and outlook}

We have presented an implementation of an extended frequentist method to estimate the systematic
uncertainties in the determination of $a_\mu^{\rm HLO}$. We consider a large number of variations
including cuts on the set of available ensembles and various fit ans\"atze to describe the momentum
dependence, lattice artifacts and the pion mass dependence. In order to deal with the large
statistical correlations among $Q^2$-points, we generate stochastic samples consisting of 30 and 40
points only. Our conservative estimate of the systematic errors dominates the overall uncertainty of
our preliminary results.\newline 
We are currently investigating various approaches to improve the accuracy of our determination of
$a_\mu^{\rm HLO}$. These include a dedicated study of the low $Q^2$ regime \cite{Golterman:2014ksa} in
combination with the use of time moments \cite{Chakraborty:2014zma}. Furthermore, we are also
investigating the mixed-representation method \cite{Francis:2013qna}, as well as using the Adler
function to compute $a_\mu^{\rm HLO}$ \cite{DellaMorte:2014rta}.\newline

\textsc{Acknowledgements}: Our calculations were performed on the ``Wilson'' and
 ``Clover'' HPC Clusters at the Institute for Nuclear Physics, University of Mainz. We thank Dalibor
 Djukanovic and Christian Seiwerth for technical support. This research has been supported in part
 by the DFG via the SFB~1044. G.H. acknowledges support by the the Spanish MINECO through the
 Ram\'on y Cajal Programme and through the project FPA2012-31686 and by the Centro de excelencia
 Severo Ochoa Program SEV-2012-0249. This work was granted access to the HPC resources of the Gauss
 Center for Supercomputing at Forschungzentrum J\"ulich, Germany, made available within the
 Distributed European Computing Initiative by the PRACE-2IP, receiving funding from the European
 Community's Seventh Framework Programme (FP7/2007-2013) under grant agreement RI-283493 (project
 PRA039) and ERC grant agreement No 279757.

\end{document}